\documentclass{article}

\usepackage{microtype}
\usepackage{graphicx}
\usepackage{subfigure}
\usepackage{booktabs} 

\usepackage[inline]{enumitem}
\usepackage{hyperref}

\usepackage[accepted]{icml2022}

\usepackage{amsmath}
\usepackage{amssymb}
\usepackage{mathtools}
\usepackage{amsthm}

\usepackage[capitalize,noabbrev]{cleveref}

\theoremstyle{plain}

\theoremstyle{definition}

\theoremstyle{remark}

\usepackage[textsize=tiny]{todonotes}

\usepackage[acronym, shortcuts, nohypertypes={acronym}]{glossaries}
\newacronym{CCC}{CCC}{concordance correlation coefficient}
\newacronym{CNN}{CNN}{convolutional neural network}
\newacronym{EXVO}{\textsc{ExVo}}{ICML Expressive Vocalisations Workshop \& Competition 2022}
\newacronym{HVB}{H-VB}{The Hume Vocal Burst Competition Dataset}
\newacronym{FFNN}{FFNN}{feed-forward neural network}
\newacronym{SER}{SER}{speech emotion recognition}
\newacronym{SGD}{SGD}{stochastic gradient descent}

\newcommand{\ie}{i.\,e.}

\newcommand{\exvo}{{\textsc{ExVo-FEW-SHOT}}}

\icmltitlerunning{Few-shot personalisation with speaker enrolment}

\begin{document}

\twocolumn[
\icmltitle{Exploring speaker enrolment for few-shot personalisation\\in emotional vocalisation prediction}



\icmlsetsymbol{equal}{*}

\begin{icmlauthorlist}
\icmlauthor{Andreas Triantafyllopoulos}{yyy}
\icmlauthor{Meishu Song}{yyy,zzz}
\icmlauthor{Zijiang Yang}{yyy}
\icmlauthor{Xin Jing}{yyy}
\icmlauthor{Bj\"{o}rn W. Schuller}{yyy,xxx}
\end{icmlauthorlist}

\icmlaffiliation{yyy}{EIHW, University of Augsburg, Augsburg, Germany}
\icmlaffiliation{zzz}{Educational Physiology Laboratory, University of Tokyo, Japan}
\icmlaffiliation{xxx}{GLAM, Imperial College, London, United Kingdom}

\icmlcorrespondingauthor{A.T.}{andreas.triantafyllopoulos@uni-a.de}

\icmlkeywords{ExVo, Few-Shot Personalisation}

\vskip 0.3in
]



\printAffiliationsAndNotice{} 

\begin{abstract}
In this work, we explore a novel few-shot personalisation architecture for emotional vocalisation prediction.
The core contribution is an `enrolment' encoder which utilises two unlabelled samples of the target speaker to adjust the output of the emotion encoder; the adjustment is based on dot-product attention, thus effectively functioning as a form of `soft' feature selection.
The emotion and enrolment encoders are based on two standard audio architectures: CNN14 and CNN10.
The two encoders are further guided to forget or learn auxiliary emotion and/or speaker information.
Our best approach achieves a CCC of $.650$ on the ExVo Few-Shot dev set, a $2.5\%$ increase over our baseline CNN14 CCC of $.634$.
\end{abstract}

\section{Introduction}
\label{sec:intro}

The modelling of emotional expression through non-verbal vocalisations is recently undergoing a paradigm shift following the introduction of a new coding scheme that imposes a continuous scale over several emotion categories~\citep{Cowen19-BHV}.
Research findings suggest that humans are better able to perceive positive emotions from non-verbal vocalisations than prosody~\citep{Hawk09-WORTH, Sauter13-CRE} and emerging \ac{SER} approaches utilise that information to improve \ac{SER} performance~\citep{Huang19-NVS, Hsu21-NVA}.
To that end, the \ac{EXVO}~\citep{Baird22-EXVO} presents an excellent opportunity for studying the automatic recognition of emotion through vocalisations.

The present contribution targets the {\exvo} task, whose goal is to study the efficacy of personalisation techniques using two-shot adaptation.
Emotion researchers have long investigated individualistic effects in emotional expressivity, which is attributable to differences in personality, childhood, and context~\citep{Larsen87-IDC, Barr09-IDEE}.
Accordingly, several prior works have attempted to incorporate those differences in the automatic recognition of emotion~\citep{Rahman12-FAS, Rudovic18-PML, Yin20-SIA, Li20-SI, Triantafyllopoulos21-DSC, Moine21-SASER, Sridhar22-UPSER, Fan22-ISNet}.
Our approach is mostly inspired from \citet{Li20-SI, Triantafyllopoulos21-DSC, Moine21-SASER, Fan22-ISNet}.
In particular, like \citep{Triantafyllopoulos21-DSC, Moine21-SASER, Fan22-ISNet}, we use an auxiliary network that conditions a main classification network based on two enrolment utterances from the same speaker; in that sense, we are most similar to \citep{Triantafyllopoulos21-DSC, Fan22-ISNet}, as \citet{Moine21-SASER} extracts the conditioning information from the same utterance while those two 
get it from a neutral enrolment utterance.
In our case, however, the enrolment utterance is not neutral (and thus does not represent a speaker `baseline' as per \citep{Triantafyllopoulos21-DSC, Fan22-ISNet}) but contains itself some emotion (that we attempt to remove using adversarial adaptation~\citep{Ganin16-DANN}).
Furthermore, we draw from \citet{Yin20-SIA, Li20-SI}, which induce invariance to speaker characteristics using auxiliary adversarial subnetworks.
Thus, our main contribution can be summarised as follows: we explore the use of enrolment-based conditioning for few-shot personalisation using auxiliary guiding losses.
Our methodology is described in detail in \cref{sec:method}\footnote{Code: https://github.com/ATriantafyllopoulos/exvo-eihw-personalisation}.
This is followed by our results in \cref{sec:results} and a conclusion in \cref{sec:conclusion}.

\section{Methodology}
\label{sec:method}

Our work is targeted towards the {\exvo} task introduced in the \acl{EXVO}~\citep{Baird22-EXVO}.
{\exvo} is based on the \ac{HVB}~\citep{Cowen19-BHV}, a novel dataset of emotional vocalisations.
It contains over $36$ hours of data from $1702$ speakers from four countries crowdsourced in `in-the-wild' conditions.
Each burst was annotated for the ten different emotions of \emph{Amusement}, \emph{Awe}, \emph{Awkwardness}, \emph{Distress}, \emph{Excitement}, \emph{Fear}, \emph{Horror}, \emph{Sadness}, \emph{Surprise}, and \emph{Triumph}, each annotated on a scale of $[1-100]$ for each burst; thus, the main goal of \ac{EXVO} is multitask regression for all ten targets.
Performance is evaluated by taking the average \ac{CCC} over each category as a holistic score, $\hat{C}$.
The goal of {\exvo} is to improve this score by providing $2$ annotated bursts per speaker in the test set.
These bursts serve as `enrolment' utterances and enable adaptation to the individual characteristics of each speaker.

\begin{figure*}[t]
    \centering
    \includegraphics[width=\textwidth]{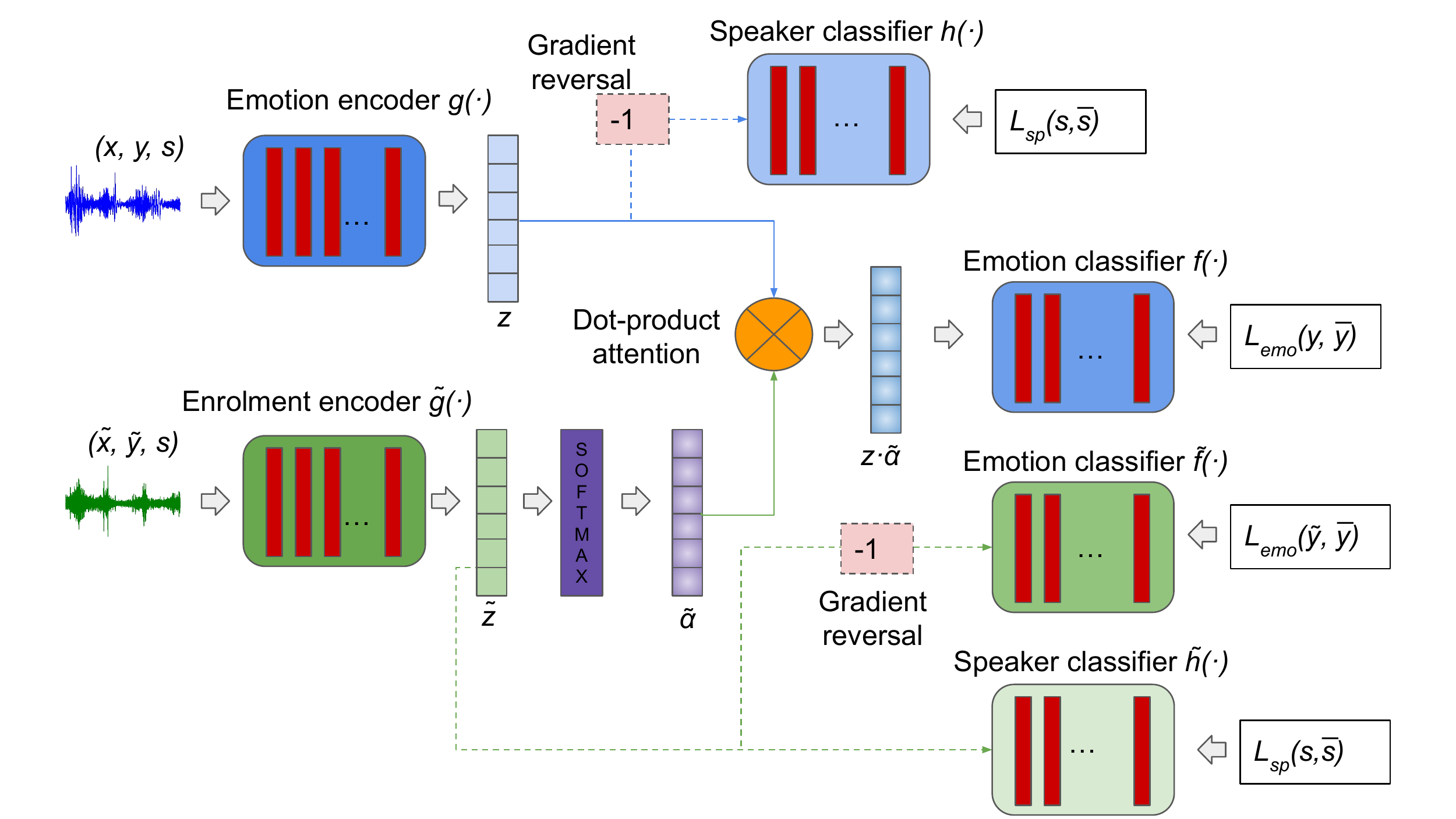}
    \caption{
    Overview of the proposed architecture for enrolment-based personalisation.
    The architecture frontend consists of two jointly-trained branches: the emotion encoder branch (top, blue) which learns emotion-specific information about the target utterance and the enrolment encoder (bottom, green) which conditions the emotion encoder on speaker information.
    Conditioning is achieved by attention; the fused output is fed into an output emotion classifier; auxiliary losses (dotted lines) are optionally introduced to strengthen speaker information in the enrolment encoder and remove speaker/emotional content from the emotion/enrolment encoder.
    }
    \label{fig:arch}
\end{figure*}

An overview of our proposed methodology is presented in \cref{fig:arch}.
Our architecture consists of three core constituents: an emotion encoder, $g(\cdot)$, an enrolment encoder, $\tilde{g}(\cdot)$, and an emotion classifier, $f(\cdot)$.
Several variants are explored by additionally utilising a set of optional components during the training 
phase: 
an adversarial speaker classifier, $h(\cdot)$, attached to the emotion encoder, as well as an adversarial emotion classifier, $\tilde{f}(\cdot)$, and a speaker classifier $\tilde{f}(\cdot)$ attached to the enrolment encoder.
The rationale and function of each component is explained below.

The emotion encoder $g(\cdot)$ is tasked with extracting emotional information from input vocalisations; its input $x$ (with label $y$ obtained from speaker $s$) is the target vocalisation on which the model is trained/evaluated.
The encoder consists of several differentiable layers which generate an emotional embedding $z$.
In the baseline variant of our approach, this emotional embedding is fed into the emotion classifier $f(\cdot)$, which generates the emotion estimates $\bar{y}$ -- these are then used to train the system by forming the input to the loss function $L_{emo}(y, \bar{y})$.

Our core contribution is the introduction of an enrolment encoder $\tilde{g}(\cdot)$, which is tasked with \emph{adapting} (personalising) to the target speaker $s$.
It accepts as input a small 
set of exemplar enrolment utterances $\tilde{x}$ (with label $\tilde{y}$) from speaker $s$ (the same speaker who generated the target utterance $x$).
The encoder consists of several differentiable layers which generate an adaptation embedding $\tilde{z}$.
Ideally, this embedding should encode information about how the target speaker is expressing their emotions through vocalisations.
In this work, we use $\tilde{z}$ to impose a soft feature selection process on $z$ through the use of \emph{dot-product attention}~\citep{Bahdanau15-NMT}: $\tilde{z}$ is first passed through a softmax layer to obtain $\tilde{\alpha}$ which is then multiplied with $z$ ($z\cdot\tilde{\alpha}$) to highlight the parts of $z$ that are most relevant for this particular speaker.
In order to stabilise the performance of the dot-product attention, we introduce a residual connection, so the input to the emotion classifier is $e = z + z\cdot\tilde{\alpha}$ -- as is often done in other works using attention~\citep{Vaswani17-AYN}.

The other three components of our architecture are optional and are introduced in an attempt to help the emotion and enrolment encoders learn their intended functions: the emotion encoder should learn a speaker-agnostic embedding $z$ that the enrolment encoder will help personalise through a speaker-aware embedding $\tilde{z}$.
This conceptualisation offers the following design principle: we should guide the emotion encoder to `forget' any speaker information and help the enrolment encoder learn more of it.
This is achieved by the additional speaker classifiers, $h(\cdot)$ and $\tilde{h}(\cdot)$, respectively.
Both accept as input the corresponding embeddings $z$ and $\tilde{z}$ and output a predicted speaker label $\bar{s}$; they are then trained with a speaker loss $L_{sp}(c, \bar{s})$.
However, there is one crucial difference: the speaker classifier attached to the emotion encoder is preceded by a \emph{gradient reversal layer}.
This layer was introduced by \citet{Ganin16-DANN} for domain adaptation, and has been widely used to remove undesired information from learnt representations -- including for speaker-invariant \ac{SER}~\citep{Li20-SI}.
The functionality of this layer is simple: it merely inverts the gradients forcing the emotion encoder to perform \emph{gradient ascent} (as it continues to minimise the error which now corresponds to the inverse of the speaker loss).

Finally, the last optional component of our architecture is an (adversarial) emotion classifier, $\tilde{f}(\cdot)$, attached to the output of the enrolment encoder.
This is introduced to solve a further challenge imposed by {\exvo}: unlike other works which relied on enrolment utterances lacking emotion (\ie, neutral ones)~\citep{Triantafyllopoulos21-DSC, Fan22-ISNet}, the utterances provided here are themselves emotional and using them to adapt to the target speaker might inadvertently remove relevant information.
Thus, we hypothesise that a gradient reversal layer which removes emotional information from the enrolment encoder will be beneficial for generalisation to new speakers.

All-in-all we explore:
\begin{enumerate}[noitemsep,nolistsep]
    \item $g \circ f$ (baseline), which uses only the emotion encoder and classifier,
    \item $g \circ f \circ (-h)$, which forces the emotion encoder to forget speaker information but makes no use of enrolment,
    \item $g \circ f \circ \tilde{g}$, which additionally introduces enrolment to the baseline,
    \item $g \circ f \circ \tilde{g} \circ \tilde{h}$, which guides the enrolment encoder with speaker information,
    \item $g \circ f \circ \tilde{g} \circ (-\tilde{f})$, which guides the enrolment encoder away from emotional information,
    \item $g \circ f \circ \tilde{g} \circ \tilde{h} \circ (-\tilde{f})$, a combination of 4) and 5)
    \item $g \circ f \circ \tilde{g} \circ (-h) \circ \tilde{h} \circ (-\tilde{f})$, a combination of 2), 4), and 5).
\end{enumerate}
We note that none of the approaches investigated here requires \emph{annotated} emotional information during test time; it is merely enough to provide $2$ unlabelled enrolment utterances of the target speaker.

Given its multiple components, our architecture comes with several knobs to twist.
In the present work, we limit ourselves to fairly generic hyperparameters and leave a more thorough ablation study for follow-up work.
Thus, for the learnable components, we rely on standard auditory architectures.
The emotion encoder is identical to (the convolution part of) CNN14 and the enrolment one to CNN10, both introduced in \citet{Kong19-PANNS}.
Both these architectures have proven successful in a variety of tasks -- including \ac{SER}~\citep{Triantafyllopoulos21-RTA}.
The classifiers are accordingly identical to the output parts of CNN14 and CNN10.
Specifically, $f(\cdot)$ and $h(\cdot)$ are $2$-layered \acp{FFNN} each with $2048$ hidden units and an appropriate output dimension, while $\tilde{f}(\cdot)$ and $\tilde{h}(\cdot)$ are also $2$-layered \acp{FFNN} with $512$ hidden units.

All models are trained for $120$ epochs.
We used a batch size of $8$, a \ac{SGD} optimiser with Nesterov momentum of $0.9$, and a starting learning rate of $0.001$, which is reduced by a factor of $0.1$ whenever the development set score stops decreasing for $5$ epochs.
The best model is selected based on the development set performance, which is evaluated at the end of each epoch.
$L_{sp}(\cdot, \cdot)$ is set to the standard cross-entropy loss; $L_{emo}(\cdot, \cdot)$ to the \ac{CCC} loss commonly used in dimensional \ac{SER} tasks -- including the ExVo baseline~\citep{Baird22-EXVO}.
Whenever multiple classifiers are used, their losses are simply averaged to procure the final loss with which the entire architecture is optimised (finding a more appropriate combination of weights for all losses is another knob that could be tweaked to further improve performance).
Moreover, for each instance in the training, the enrolment encoder is fed with $2$ utterances from the same speaker selected randomly; this randomisation is there to improve generalisation to test set conditions where the enrolment utterances are \textit{a priori} unknown.
To make our results reproducible, 
the development set results are obtained by selecting the first $2$ utterances per development set speaker as they appear in the \ac{HVB} data.
Finally, as bursts can be of different length, we randomly crop both the target and the enrolment utterances  to $2.5$\,s (average burst duration is $2.23$\,s; longer utterances are randomly cropped; silence is randomly added to the start and/or end times of shorter utterances).
During evaluation, we use a batch size of $1$, so the target utterances can be fed in whole, while the shorter of the enrolment utterances is zero-padded to match the duration of the longest one.
\begin{table*}[t]
    \centering
    \caption{
    Ablation study for the different architecture components ($g$: emotion encoder; $\tilde{g}$ enrolment encoder; $f/h$: emotion/speaker classifier on the main branch; $\tilde{f}/\tilde{h}$: emotion/speaker classifier on the enrolment branch; $(-\cdot)$: module preceded by a gradient reversal layer).
    We show the official {\exvo} evaluation metric ($\hat{C}$: mean \ac{CCC} over all emotions) with 95\,\% CIs on the development set.
    Additionally, we show test set performance for our top-5 results.
    Finally, we show the relative gain over the baseline ($g \circ f$).
    }
    \label{tab:results}
    \begin{tabular}{l|rr|rr}
        \toprule
        & \multicolumn{2}{c|}{\textbf{Development set}} & \multicolumn{2}{c}{\textbf{Test set}}\\
        \textbf{Architecture} & \multicolumn{1}{c}{\textbf{$\hat{C}$ [CI]}} & \multicolumn{1}{c|}{\textbf{\% Gain}} & \multicolumn{1}{c}{\textbf{$\hat{C}$}} & \multicolumn{1}{c}{\textbf{\% Gain}}\\
        \midrule
        {\exvo} Baseline & N/A & N/A & .444 & N/A\\
        \midrule
        $g \circ f$ & .645 [.640-.650] & N/A & .634 & N/A\\
        $g \circ f \circ (-h)$ & .608 [.603-.613] & -5.7 & N/A & N/A\\
        $g \circ f \circ \tilde{g}$ & .656 [.651-.661] & +1.7 & \textbf{.650} & \textbf{+2.5}\\
        $g \circ f \circ \tilde{g} \circ \tilde{h}$ & .655 [.649-.660] & +1.6 & .639 & +0.8\\
        $g \circ f \circ \tilde{g} \circ (-\tilde{f})$ & \textbf{.658 [.653-.663]} & \textbf{+2.0} & .647 & +2.0\\
        $g \circ f \circ \tilde{g} \circ \tilde{h} \circ (-\tilde{f})$ & .652 [.646-.657] & +1.1 & .642 & +1.3\\
        $g \circ f \circ \tilde{g} \circ (-h) \circ \tilde{h} \circ (-\tilde{f})$ & .636 [.630-.641] & -1.4 & N/A & N/A\\
        \bottomrule
    \end{tabular}
\end{table*}
The gradient reversal layer requires additional tuning~\citep{Ganin16-DANN}.
Specifically, like \citet{Ganin16-DANN}, we gradually introduce gradient reversal in our training using a warm-up schedule.
For the first $10$ epochs of training, the $\lambda$ hyperparameter is set to $-1$,
while from epoch $10$ to $60$ we linearly increase $\lambda$ to $1$.

\section{Results}
\label{sec:results}


In \cref{tab:results}, we present development set results for all investigated alternatives, as well as test set results for our top-5 architectures as found on the development set. 
In addition to the standard score metric of {\exvo}, the average \ac{CCC} of all ten emotions ($\hat{C}$), we also show $95\%$ confidence intervals (CIs) for the development set, obtained via $1000$-sampled bootstrapping (with replacement)
and relative gain over our baseline.
Our first observation is that our baseline, which comprises solely of CNN14, vastly outperforms the official {\exvo} baseline, which relied on 1D-CLSTM networks trained on raw audio inputs~\citep{Tzirakis17-E2E}, with a $\hat{C}$ of $.634$ on the test set, compared to a baseline of $.444$.
This indicates that 2D-CNN architectures based on log-Mel spectrograms might be more suitable for the analysis of emotional vocalisations.
Interestingly, inducing speaker invariance by adding an auxiliary adversarial speaker classification output to the main emotion encoder branch ($g \circ f \circ (-h)$) substantially reduces development set performance.

Adding enrolment-based personalisation yields small improvements, though in most cases the development set CIs do not overlap with those of our baseline, lending some importance to them.
The highest improvement is obtained by adding the enrolment encoder without any classifiers attached to it ($g \circ f \circ \tilde{g}$); this results in a $+2.5\,\%$ increase to a $\hat{C}$ of $.650$.
This is slightly better than adding an adversarial emotion classifier to the enrolment branch, which yields a $\hat{C}$ of $.647$, followed by adding a speaker classification decoder ($\hat{C}$ of $.639$) and using both decoders ($\hat{C}$ of $.642$).
Further adding an adversarial speaker classifier in the main encoder branch ($g \circ f \circ \tilde{g} \circ \tilde{h}$) actually decreases development set performance ($\hat{C}$ of $.636$) -- which is attributable to the negative effect observed by adding $(-h)$ to the main branch.

Overall, it seems that most of the gain is obtained by adding the enrolment encoder, with the auxiliary outputs that guide it to learn more speaker information and forget emotional information adding marginal improvements over that on the development that do not translate well to the test set.
This indicates that the additional information injected to the embeddings of the emotion encoder is sufficient for adapting to the new speaker.
The relative gain of $+2.0\,\%$ is comparable with that reported in previous work: for example, \citet{Sridhar22-UPSER} report a $+13.5\,\%$ for valence, but only $+1.8\,\%$ for arousal and $+1.2\,\%$ for dominance, while \citet{Li20-SI} and \citet{Fan22-ISNet} report a $+4.2\,\%$\,/\,$3.3\,\%$ gain for categorical emotion recognition, respectively.



\section{Conclusion}
\label{sec:conclusion}

We presented a novel, few-shot personalisation architecture for emotional vocalisation prediction.
Our results show that the proposed approach can overcome a standard adversarial loss which induces speaker invariance.
We included a comprehensive ablation study of the different components of our architecture and conclude that most of the improvement stems from the enrolment encoder.
Given that our model consists of multiple parts, further work is needed to tune them and improve their combination.
\section{Acknowledgements}
This research was partially supported by the Deutsche Forschungsgemeinschaft (DFG) under grant agreement No.\ 442218748 (AUDI0NOMOUS).

\bibliography{example_paper}
\bibliographystyle{icml2022}

\end{document}